\documentclass[a4,12pt]{article}
\usepackage{latexsym}
\usepackage{comment}
\usepackage[pdftex]{graphicx}
\usepackage{amsmath,amssymb}
\usepackage{amscd,amsmath,amssymb,amsfonts,xspace,mathrsfs}

\usepackage{epsf}

\def\varpi{t}

\def\Im{\,{\rm Im}\,}

\def\({\left(}
\def\){\right)}
\def\[{\left[}
\def\]{\right]}
\def\<{\left\langle}
\def\>{\right\rangle}
\def\hf{{1\over 2}}

\newcommand{\CP}{\IC P^1}

\newcommand{\bfk}{{\boldsymbol k}}

\renewcommand{\d}{\mathrm{d}}
\newcommand{\de}{\mathrm{d}}

\newcommand{\I}{\mathrm{i}}

\newcommand{\p}{\partial}

\newcommand{\cV}{\mathcal{V}}

\newcommand{\cK}{\mathcal{K}}
\newcommand{\cM}{\mathcal{M}}

\newcommand{\cN}{\mathcal{N}}
\newcommand{\cE}{\mathcal{E}}
\newcommand{\cX}{\mathcal{X}}

\newcommand{\cR}{\mathcal{R}}

\newcommand{\cJ}{\mathcal{J}}
\newcommand{\cZ}{\mathcal{Z}}

\newcommand{\cA}{\mathcal{A}}
\newcommand{\cB}{\mathcal{B}}

\newcommand{\cY}{\mathcal{Y}}

\DeclareSymbolFont{AMSa}{U}{msa}{m}{n}
\DeclareSymbolFont{AMSb}{U}{msb}{m}{n}
\DeclareMathSymbol{\fieldR}{\mathalpha}{AMSb}{"52}

\newcommand{\N}{{\mathcal N}}

\newcommand{\nn}{\nonumber}

\newcommand{\IR}{\mathbb{R}}
\newcommand{\IC}{\mathbb{C}}
\newcommand{\IZ}{\mathbb{Z}}

\newcommand{\tzeta}{\tilde\zeta}

\newcommand{\thh}{\tilde h}
\newcommand{\tlk}{\tilde k}

\def\bea{\begin{eqnarray}}
\def\eea{\end{eqnarray}}
\def\be{\begin{equation}}
\def\ee{\end{equation}}
\def\ba{\begin{align}}
\def\ea{\end{align}}
\def\bse{\begin{subequations}}
\def\ese{\end{subequations}}

\def\bX{\bar X}
\def\bF{\bar F}

\def\bZ{\bar Z}
\def\bE{\bar E}

\def\ba{\bar a}

\def\bi{\bar \imath}
\def\bj{\bar \jmath}

\def\bz{\bar z}

\def\hN{\hat N}

\def\ci#1{c^{[#1]}}
\def\cij#1{c^{[#1]}}

\newcommand{\Li}{{\rm Li}}

\def\ellg#1{\ell_{#1}}

\def\nk{n_{k}^{(0)}}

\def\hng#1{\Omega_{\gamma_{#1}}}
\def\Om#1{\Omega_{#1}}

\def\Ilg{\cJ^{(1)}}
\def\Ilog#1{\cJ^{(1,#1)}}
\def\Irt{\cJ^{(2)}}
\def\Irat#1{\cJ^{(2,#1)}}

\def\Igp{\Ilog{+}_{\gamma}}
\def\Igm{\Ilog{-}_{\gamma}}
\def\Igpm{\Ilog{\pm}_{\gamma}}

\def\Igg#1{\Ilg_{\gamma_{#1}}}

\def\rIg{\Irt_{\gamma}}
\def\rIgp{\Irat{+}_{\gamma}}
\def\rIgm{\Irat{-}_{\gamma}}
\def\rIgpm{\Irat{\pm}_{\gamma}}

\def\cij#1{c}
\def\ci#1{c}

\def\vl{v}
\def\bvl{\bar \vl}

\def\Min{M}

\def\Uin{\mathbf{U}}

\def\cVs{\cV_{(\sigma)}}

\def\cMsk{\cM_{\rm sk}}
\def\CY{\mathfrak{Y}}

\def\Fcl{F^{\rm cl}}
\def\bFcl{\bF^{\rm cl}}

\def\Va{V^{(\varphi)}}

\def\caja{\mathsurround=0pt}
\def\eqalign#1{\,\vcenter{\openup2\jot \caja
        \ialign{\strut \hfil$\displaystyle{##}$&$
        \displaystyle{{}##}$\hfil\crcr#1\crcr}}\,}

\begin{document}
\thispagestyle{empty}

{\hbox to\hsize{
\vbox{\noindent IPMU17-0029 \hfill June 2017 \\
NORDITA-2017-23 \hfill revised version}}}

\noindent
\vskip2.0cm
\begin{center}

{\large\bf No inflation in type IIA strings on rigid CY spaces}
\vglue.3in

Yuki Wakimoto~${}^{a}$ and Sergei V. Ketov~${}^{a,b,c,d}$  
\vglue.1in
${}^a$~Department of Physics, Tokyo Metropolitan University, \\
Minami-ohsawa 1-1, Hachioji-shi, Tokyo 192-0397, Japan \\
${}^b$~Institute of Physics and Technology, Tomsk Polytechnic University,\\
30 Lenin Avenue, Tomsk 634050, Russian Federation \\
${}^c$ Nordita, KTH Royal Institute of Technology and Stockholm University,
Roslagstullsbacken 23, SE-106 91 Stockholm, Sweden\\
${}^d$~Kavli Institute for the Physics and Mathematics of the Universe (IPMU),
\\The University of Tokyo, Chiba 277-8568, Japan \\
\vglue.1in
wakimoto-yuki@ed.tmu.ac.jp, ketov@tmu.ac.jp,
\end{center}

\vglue.3in

\begin{center}
{\Large\bf Abstract} 
\end{center}

\noindent We investigate whether cosmological inflation is possible in a class of flux compactifications of type IIA strings on rigid Calabi-Yau manifolds, when all perturbative string corrections are taken into account.  We confine ourselves to the universal hypermultiplet and an abelian vector multiplet, representing matter in four dimensions. Since all axions can be stabilized by D-instantons, we choose dilaton and a K\"ahler modulus as the only running scalars. Though positivity of their scalar potential can be achieved, we find that there is no slow roll ($\varepsilon > 13/6$), and no Graceful Exit because the scalar potential has the run-away behaviour resulting in decompactification. We conclude that it is impossible to generate phenomenologically viable inflation in the given class of flux compactifications of type IIA strings, without explicit breaking of $N=2$ local supersymmetry of the low-energy effective action.

\newpage

\section{Introduction}
\label{sec-intro}

The main problem of phenomenological applications of string theory to cosmology and particle physics is moduli stabilization
\cite{Baumann:2014nda}. On the one hand, though it is possible to stabilize all moduli at the classical level \cite{DeWolfe:2005uu}, the existing no-go theorems forbid de Sitter (dS) vacua in the classical supergravity compactifications \cite{Maldacena:2000mw,Ivanov:2000fg} and, hence, require taking into account either quantum corrections or non-geometric fluxes \cite{Kachru:2003aw,Balasubramanian:2005zx,Conlon:2005ki,Westphal:2006tn,deCarlos:2009qm,Louis:2012nb,Danielsson:2012by,Blaback:2013qza,Hassler:2014mla}.
On the other hand, all phenomenologically viable inflationary models are sensitive to Planck-scale physics and thus require
a UV-completion, which raises the question about  derivation of any of them from a fundamental theory of quantum gravity such as superstrings. To the best of our knowledge, there is no compelling inflationary model of string cosmology, despite of huge theoretical efforts and many proposals in the literature \cite{Baumann:2014nda}.

As regards the type IIA strings compactified on Calabi-Yau (CY) threefolds, there is the "no-go" statement (i.e. no possibility of  inflation) in the literature  \cite{Hertzberg:2007wc}, derived in the certain (semi-classical)  limit at large volume and small string coupling, by using the scaling arguments. However, it is worthwhile to raise the same question {\it beyond} the semi-classical limit (i.e. for any values of volume and string coupling), by including (quantum) string corrections both in $\alpha'$ and $g_s$, thus also beyond the ten-dimensional IIA supergravity approximation, where the assumptions used in \cite{Hertzberg:2007wc} do not apply.

Quantum corrections are under control in the case of type II string compactifications on CY. In this case, the low energy effective action (LEEA) in four dimensions preserves $N=2$ local supersymmetry (8 supercharges) and is completely determined by the geometry of its moduli space spanned by the scalar fields of $N=2$ vector- and hyper-multiplets. The only way to generate a scalar potential in the effective $N=2$ supergravity in four dimensions is via adding the NS- and RR-fluxes leading to the gauging of some of the isometries of the moduli space of the original fluxless compactification.  Actually, the integrated Bianchi identities give rise to certain tadpole cancellation conditions, which in the presence of fluxes generically can be satisfied only by adding orientifolds reducing supersymmetry to $N=1$ \cite{Giddings:2001yu}. However, in type IIA string theory it is possible to choose such fluxes that the tadpole cancellation condition holds automatically as e.g., is the case with the NS $H$-fluxes and the RR $F_4$- and $F_6$-fluxes, provided that one ignores their backreaction \cite{Kachru:2004jr}. We follow the same strategy in this paper.

Explicit calculations are possible in the case of a {\it rigid} CY threefold $\CY$. Such manifold has the vanishing Hodge number $h^{2,1}(\CY)=0$, so that the LEEA is described by $N=2$ supergravity interacting with the so-called {\it universal hypermultiplet} (UH), and some number $h^{1,1}(\CY)> 0$ of vector multiplets. As was found in \cite{Alexandrov:2016plh}, this class of $N=2$ compactifications does not allow meta-stable vacua, within validity of the approximation used. However, it is still has to be checked whether the same scalar potential is suitable for slow roll inflation. In this paper, we restrict ourselves to the perturbative approximation where all instanton contributions are neglected, but the perturbative $\alpha'$ and $g_s$-corrections are retained. For even more simplicity, we consider the case with only one K\"ahler modulus, i.e. a rigid CY with $h^{1,1}=1$.  Since no such rigid CY space was found yet, our investigation should be merely viewed as a negative test (i.e. no formal proof of our title yet). The use of fictitious CY is commonplace in modern string theory. 

Our paper is organized as follows. In the next section 2 we review basic information about the relevant CY string compactifications with fluxes, and provide the scalar potential in the corresponding (gauged) $N=2$ supergravity by following 
\cite{Alexandrov:2016plh} that is our starting point. In section \ref{sec-pert} we study the perturbative scalar potential and
slow roll conditions. Sec.~4 is our conclusion. Our notation about special and quaternionic geometries is collected in Appendix A. Details of the  UH moduli space metric are collected in Appendix B.

\section{Scalar potential in type-II flux-compactifications}
\label{sec-potential}

In this section we recall the known facts about the scalar potential in {\it generic} type II string compactifications with fluxes, in the context of $N=2$ supergravity in four dimensions by following \cite{Alexandrov:2016plh}, and describe our setup.

The four-dimensional LEEA of type II strings compactified on a Calabi-Yau threefold $\CY$
is given by $N=2$ supergravity coupled to $N=2$ vector and hypermultiplets.
In the two-derivative approximation, where one ignores the higher curvature terms appearing as $\alpha'$-corrections,
the bosonic part of the action comprises only kinetic terms for the metric, vector and scalar fields
arising after compactification. The couplings of these kinetic terms are non-trivial, being
restricted by $N=2$ supersymmetry in terms of the metrics on the vector and hypermultiplet moduli spaces,
$\cM_V$ and $\cM_H$, parametrized by the scalars of the corresponding multiplets.
Furthermore, $N=2$ supersymmetry restricts $\cM_V$ to be a special K\"ahler manifold, with a K\"ahler potential $\cK(z^i,\bz^{\bi})$
(with $i=1,\dots, h^{1,1}$ in type IIA) determined by a holomorphic prepotential $F(X^I)$
(with $I=(0,i)=0,\dots ,h^{1,1}$ and $z^i=X^i/X^0$),
a homogeneous function of degree 2. Similarly, $\cM_H$ must be a quaternion-K\"ahler (QK) manifold of dimension
$4(h^{2,1}+1)$ \cite{Bagger:1983tt}.
We denote the metrics on the two moduli spaces by $\cK_{i\bj}$ and $g_{uv}$, respectively.

The resulting theory is not viable from the phenomenological point of view since
it does not have a scalar potential, so that all moduli remain unspecified.
This gives rise to the problem of {\it moduli stabilization}, i.e. generating a potential for the moduli.
This requires $N=2$ {\it gauged} supergravity. The latter can be constructed from the ungauged supergravity when 
the moduli space $\cM_V\times \cM_H$ has some isometries, which are to be gauged with respect to the vector fields $A^I$ comprising,
besides those of vector multiplets, the gravi-photon $A^0$ of the gravitational multiplet. Physically, this means that the scalar fields affected by the isometries acquire charges under the vector fields used in the gauging.
The charges are proportional to the components of the Killing vectors $k_\alpha$ corresponding to the gauged isometries.
We consider only abelian gaugings of isometries of the hypermultiplet moduli space $\cM_H$ because
quantum corrections are known to break any non-abelian isometries. Then the charges are characterized by the vectors
$\bfk_I=\Theta_I^\alpha k_\alpha\in T\cM_H$ where $\Theta_I^\alpha$ is known as the embedding tensor.

The geometry of the moduli space together with the charge vectors {\it completely} fix the scalar potential as
\cite{D'Auria:1990fj,Andrianopoli:1996cm,deWit:2001bk}~\footnote{See Appendix B and Ref.~\cite{Alexandrov:2016plh} for more details about our notation.}
\bea
V &=&
4 e^\cK \bfk^u_I \bfk^v_J g_{uv} X^I \bX^J + e^\cK \(\cK^{i\bj}D_i X^I D_{\bj}\bX^J-3 X^I \bX^J\)\(\vec\mu_I\cdot \vec \mu_J\),
\label{scpot-gen}
\eea
where $D_i X^I=(\p_i+\p_i \cK)X^I$
and $\vec\mu_I$ is the triplet of moment maps which quaternionic geometry of $\cM_H$ assigns to each isometry $\bfk_I$ \cite{MR872143}.

In string theory, $N=2$ gauged supergravity can be obtained by adding closed string fluxes to a CY compactification
(see \cite{Grana:2005jc} for a review). We take the common strategy (see, for instance,
\cite{Louis:2002ny,Giryavets:2003vd,Kachru:2004jr,DeWolfe:2005uu}) by ignoring the backreaction and assuming the compactification manifold to be CY. The LEEA of flux compactifications on CY is known to perfectly fit the framework of $N=2$ gauged supergravity \cite{Louis:2002ny}.

 In type IIA, the vector multiplet moduli space
$\cM_V$ describes the complexified K\"ahler moduli of $\CY$ parametrizing deformations of the K\"ahler structure and
the periods of the $B$-field along two-dimensional cycles, $z^i=b^i+\I t^i$.
The hypermultiplet moduli space $\cM_H$ consists of
\begin{itemize}
\item
$u^a$ --- complex structure moduli of $\CY$ ($a=1,\dots,h^{2,1}$),
\item
$\zeta^\Lambda,\tzeta_\Lambda$ --- RR-scalars given by periods of the RR 3-form potential along three-dimensional cycles of $\CY$
($\Lambda=(0,a)=0,\dots,h^{2,1}$),
\item
$\sigma$ --- NS-axion, dual to the 2-form $B$-field in four dimensions,
\item
$\phi$ --- dilaton, determining the value of the four-dimensional string coupling, $g_s^{-2}=e^{\phi}\equiv r$.
\end{itemize}

The Kaluza-Klein reduction from ten dimensions \cite{Louis:2002ny} leads to the classical metrics on $\cM_V$
and $\cM_H$. The former is the special K\"ahler metric $\cK_{i\bj}$ given by the derivatives of the K\"ahler potential
\be \label{kinpot}
\cK=-\log\[ \I\(\bX^I \Fcl_I-X^I\bFcl_I\)\],
\ee
where $\Fcl_I=\p_{X^I}\Fcl$ are the derivatives of the classical holomorphic prepotential
\be
\Fcl(X)=-\kappa_{ijk}\, \frac{X^iX^j X^k}{6X^0},
\label{Fcl}
\ee
which is determined by the triple intersection numbers $\kappa_{ijk}$ of $\CY$.
The hypermultiplet metric is given by the so-called {\it c-map} \cite{Cecotti:1989qn} which produces
a QK metric out of another holomorphic prepotential, characterizing the complex structure moduli, which carries
a Heisenberg group of continuous isometries acting by shifts on the RR-scalars and the NS-axion.
The corresponding Killing vectors are
\be
\label{heis0}
k^\Lambda=\p_{\tzeta_\Lambda}-\zeta^\Lambda\p_\sigma,
\qquad
\tlk_\Lambda=\p_{\zeta^\Lambda}+\tzeta_\Lambda\p_\sigma,
\qquad
k_\sigma=2\p_\sigma.
\ee
It is these isometries that are gauged by adding fluxes.

Type IIA strings on CY admit the NS-fluxes described by the field strength of the $B$-field:
\be
H^{\rm flux}_3=h^\Lambda\tilde\alpha_\Lambda-\thh_\Lambda\alpha^\Lambda,
\ee
where $(\alpha^\Lambda,\tilde\alpha_\Lambda)$ is a symplectic basis of harmonic 3-forms,
and the RR-fluxes given by the 2- and 4-form field strengths
\be
F^{\rm flux}_2=-m^i\tilde\omega_i,
\qquad
F^{\rm flux}_4=e_i\omega^i,
\ee
where $\tilde\omega_i$ and $\omega^i$ are bases of $H^2(\CY)$ and $H^4(\CY)$, respectively.

As regards the metric on $\cM_V$, it receives the $\alpha'$-corrections which
are all captured by a modification of the holomorphic prepotential \eqref{Fcl} as \cite{Candelas:1990rm,Hosono:1993qy} 
\be
F(X)=\Fcl(X)+\chi_\CY\,\frac{\I\zeta(3)(X^0)^2}{16\pi^3}
-\frac{\I(X^0)^2}{8\pi^3}\sum_{k_i\gamma^i\in H_2^+(\CY)}\nk \Li_3\(e^{2\pi\I k_iX^i/X^0}\),
\label{Ffull}
\ee
where $\chi_\CY=2(h^{1,1}-h^{2,1})$ is Euler characteristic of CY, $\nk$ are the genus-zero Gopakumar-Vafa invariants,
and the sum goes over the effective homology classes, i.e. $k_i\ge 0$ for all $i$, with not all of them vanishing simultaneously.
The quantum terms are given by a sum of the perturbative correction and the contribution of worldsheet instantons, respectively.  The non-perturbative quantum corrections are known to be important for stabilizing all axions
\cite{Alexandrov:2016plh}. We ignore the axions in what follows, by assigning heavy masses to them via tuning the parameters of our model.~\footnote{The alternative possibility arises when some of the axions are lighter than the other
moduli. This case is technically more involved and is not studied here.}

As regards $\cM_H$, though its complete non-perturbative description is still beyond reach, a significant progress
in this direction was achieved by using twistorial methods (see \cite{Alexandrov:2011va,Alexandrov:2013yva} for reviews).
In contrast to $\cM_V$, the hypermultiplet metric is exact in $\alpha'$, but receives $g_s$-corrections. At the perturbative level, 
it is known explicitly \cite{Alexandrov:2007ec}, and is given by a one-parameter deformation of the classical c-map metric,
whose  deformation parameter is controlled by $\chi_\CY$ \cite{Antoniadis:1997eg,Antoniadis:2003sw,RoblesLlana:2006ez}.
At the non-perturbative level, the metric gets the instanton contributions coming from D2-branes wrapping 3-cycles
(and, hence, parametrized by a charge $\gamma=(p^\Lambda, q_\Lambda)$) and NS5-branes wrapping the whole CY.
The D-instantons were incorporated to all orders using the twistor description of QK manifolds
\cite{RoblesLlana:2006is,Alexandrov:2008nk,Alexandrov:2008gh,Alexandrov:2009zh}, so that
only NS5-instanton contributions still remain unknown.

In the case of {\it rigid} CY, capital Greek indices can be omitted, so that the  $\cM_H$  has the lowest possible dimension and thus represents the simplest theoretical laboratory for explicit calculations 
\cite{Strominger:1997eb,Gutperle:2000sb,Ceresole:2001wi,Antoniadis:2003sw,Davidse:2004gg,Kachru:2004jr,Bao:2009fg,Catino:2013syn}. The metric on four-dimensional QK spaces allows an explicit parametrization
\cite{Przanowski:1991ru,MR1423177}, which reduces it to a solution of an integrable system.
In the presence of a continuous isometry, it is encoded in a solution of the integrable {\it Toda} equation
\cite{Ketov:2001ky,Ketov:2001gq,Ketov:2002vr,Davidse:2005ef,Alexandrov:2006hx,Alexandrov:2012np}.

The UH metric reads \cite{Alexandrov:2014sya}
\be
\eqalign{
\de s^2= & 
\frac{2}{r^2}\left[\(1-\frac{2r}{\cR^2\Uin}\) \((\de r)^2+\frac{\cR^2}{4}\,|\cY|^2\) \right. \cr
& \left. 
+\frac{1}{64}\(1-\frac{2r}{\cR^2\Uin}\)^{-1}\(\de \sigma +\tzeta \de \zeta-\zeta\de \tzeta+\cVs \)^2\right], \cr}
\label{mett-UHMmain}
\ee
where the functions $\cR$, $\Uin$, $\cY$ and $\cVs$, are defined in Appendix B.

The charge vectors corresponding to our choice of fluxes and generating the isometries of the metric \eqref{mett-UHMmain} 
are given by \cite{Alexandrov:2016plh}
\be
\begin{split}
\bfk_0=&\,
\thh\p_{\tzeta}+h\p_\zeta +\(2e_0+h\tzeta-\thh\zeta\) \p_\sigma\, ,
\\
\bfk_i=&\,
2e_i\p_\sigma\, .
\end{split}
\label{kilv-H}
\ee

The associated moment maps $\vec\mu_I$ are \cite{Alexandrov:2016plh}
\be
\begin{split}
\mu_i^+=&\, 0,
\qquad\qquad\qquad\quad
\mu_i^3=\frac{e_i}{2r},
\\
\mu_0^+=&\, \frac{\I\cR}{2r}\(\thh-\lambda h\),
\qquad
\mu_0^3=\frac{1}{2r}\(e_0+h\tzeta-\thh\zeta\).
\end{split}
\label{momentmap-main}
\ee

Using all the data above in Eq.~(\ref{scpot-gen})  results in the scalar potential \cite{Alexandrov:2016plh}
\be
\eqalign{
V=&\, \frac{e^{\cK}}{4r^2}\Biggl[ \frac{2|E+\cE|^2}{1-\frac{2r}{\cR^2\Uin}}
+\cK^{i\bj}\(e_i+E\cK_i\)\(e_j+\bE\cK_{\bj}\)-3|E|^2 \cr
& +4\cR^2|\thh-\lambda h|^2\(\cK^{i\bj}\cK_i\cK_{\bj}-1-\frac{4r}{\cR^2\Uin}\) 
\Biggr], \cr }
\label{potential-main}
\ee
where $\cK_i=\p_i\cK$ and 
\be
\begin{split}
E=&\, e_0+h\tzeta-\thh\zeta+e_i z^i,
\\
\cE=&\, \hf\(h\iota_{\p_\zeta}+\thh\iota_{\p_{\tzeta}}\)\cVs.
\end{split}
\label{defE}
\ee

Both the metric and the potential are invariant under
the symplectic transformations induced by a change of basis of 3-cycles on $\CY$.
This invariance can be used to put $h$-flux to zero, which we assume from now on.
In this symplectic frame, only electrically charged instantons contribute to the potential.
Using this simplification, one can show that
\be
\cE=\frac{4\thh r\bvl}{\cR (|\Min|^2+|\vl|^2)} ~~,
\ee
whereas the other quantities introduced in Appendix B can be computed explicitly as \cite{Alexandrov:2016plh}
\bea
\vl &=& 384 c\sum_{q>0}s(q) q^2 \sin(2\pi q\zeta)K_1(4\pi q\cR),
\nn\\
\Min &=& 2\lambda_2+384 c\sum_{q>0}s(q) q^2 \cos(2\pi q\zeta)K_0(4\pi q\cR),
\label{res-electric}\\
r &=& \frac{\lambda_2\cR^2}{2}-c-\frac{24c\cR}{\pi}\sum_{q>0}s(q) q \cos(2\pi q\zeta)K_1(4\pi q\cR)~.
\nn
\eea

The function $\Uin$, appearing in the potential \eqref{potential-main}, is given by \eqref{Ab-UHM}. The divisor function is defined
by
\be
s(q)\equiv\sigma_{-2}(q)=\sum_{d|n}d^{-2},
\ee
and, via Eqs.~\eqref{Omq} and \eqref{def-c}, encodes the DT invariants counting the D-instantons, with the parameter $c$.
As a result, all $g_s$-corrections affecting the scalar potential are controlled by only one topological (Euler) number.

At $h=0$ the potential explicitly depends on dilaton $r$, K\"ahler moduli $t^i$, periods $b^i$ of the
$B$-field, and the RR scalar $\zeta$, being independent of another RR scalar $\tzeta$ and the
NS-axion $\sigma$. Since the last two scalars are used for the gauging, one can redefine some of the gauge fields to absorb 
them in the effective action, where these scalars disappear from the spectrum, whereas the corresponding gauge fields become massive.

In the perturbative approximation, the potential depends on the fields $b^i$ and $\zeta$, known
as {\it axions}, only through the combination $e_i b^i-\thh\zeta$ appearing in \eqref{defE}.
The other $h^{1,1}$ independent combinations of these fields enter the potential only via instanton corrections.
This shows that the instanton corrections are {\it indispensable} for axion stabilization, and allow a simple solution 
\cite{Alexandrov:2016plh},
\be
\zeta=n/2,
\qquad
b^i=\ell^i/2.
\label{vanishsol}
\ee

Having restricted ourselves to this solution, we can simplify the scalar potential as 
\be 
\begin{split}
\Va(r,t^i)\equiv \left.V\right|_{\zeta=n/2\atop b^i=\ell^i/2}
=&\,
\frac{e^{\cK}}{4r^2}\[ \frac{4r(et)^2}{\cR^2\Min-2r}
-e^{-\cK}\hN^{ij}e_i e_j
+\frac{4\thh^2\cR^2}{e^{\cK}N_{ij}t^i t^j-1}-\frac{16\thh^2 r}{\Min} \].
\end{split}
\label{potallzero}
\ee

In what follows, we add two more simplifications by (i) neglecting all non-perturbative (instanton) corrections,
and (ii) limiting ourselves to a single K\"ahler modulus, by omitting all the lower-case latin indices too.

As was shown in \cite{Alexandrov:2014sya}, the metric \eqref{mett-UHMmain} has a curvature singularity at the hypersurface
determined by the equation $r=\hf\, \cR^2\Uin$. This singularity is an artefact of our approximation. It implies that near the singularity the metric \eqref{mett-UHMmain} and, hence, the corresponding scalar potential \eqref{potallzero} cannot be trusted. In other words, we should require that $r> r_{\rm cr}$. In the perturbative approximation one has $r_{\rm cr}=-2c$.

\section{Perturbative approximation}
\label{sec-pert}

Given a single vector multiplet of matter and, hence, a single K\"ahler modulus, we can omit all the lower case latin indices in \eqref{potallzero}
and rewrite it to the form
\begin{equation}
V(\cR(r),t)=
\frac{e^{\cK}}{4r^2}
\left(
\frac{4r(et)^2}{\cR^2M^2-2r}
-e^{-\cK}Ne^2
+\frac{4\tilde{h}\cR^2}{e^{\cK}Nt^2-1}
-\frac{16 \tilde{h}^2r}{M}
\right) \label{potallzero2}
\end{equation}
as the function of only two real variables, $r$ and $t$, representing dilaton and the imaginary part of the K\"ahler modulus, respectively.
In accordance to Appendices A and B, the functions entering \eqref{potallzero} and \eqref{potallzero2} are greatly simplified in the
perturbative approximation, where all instantons are ignored, as follows:
\begin{equation}\label{potallzeroper}
e^{-\cK}=8\cV-C,\quad \cV=\frac{\kappa t^3}{6},\quad \cR=\sqrt{\frac{2(r+c)}{\lambda_2}},\quad N=2\kappa t.
\end{equation}

The function \eqref{potallzero2} subject to the definitions \eqref{potallzeroper} is a fully explicit elementary (complicated) function that can be studied both analytically and numerically (we used Wolfram Mathematica). Its profile is given in Fig.~1. Our idea is to exploit a competition of four different terms in  \eqref{potallzero2} by varying the flux parameters, in order to compile them into a slow roll inflationary potential  for some values of $r$ and $t$.

The variables $\cR$ and $t$ have to be restricted from below, $\cR>\cR_c$ and $t>t_c$, because the potential diverges at 
\begin{equation}
\cR_c = \sqrt{\frac{2c}{\lambda_2-4\lambda_2^2}}\quad \text{and}\quad
t_c = \sqrt[3]{\frac{3C}{4\kappa}}~~.
\end{equation}

\begin{figure}[t]
\begin{center}
\includegraphics[width=6cm]{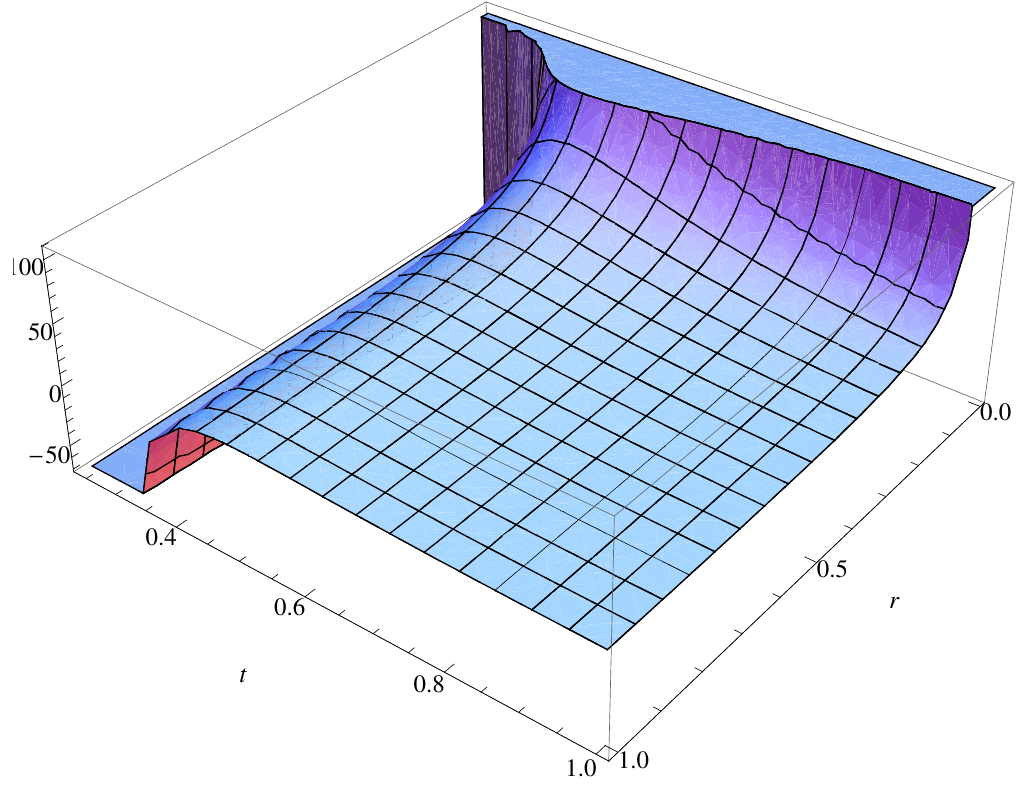}
\includegraphics[width=6cm]{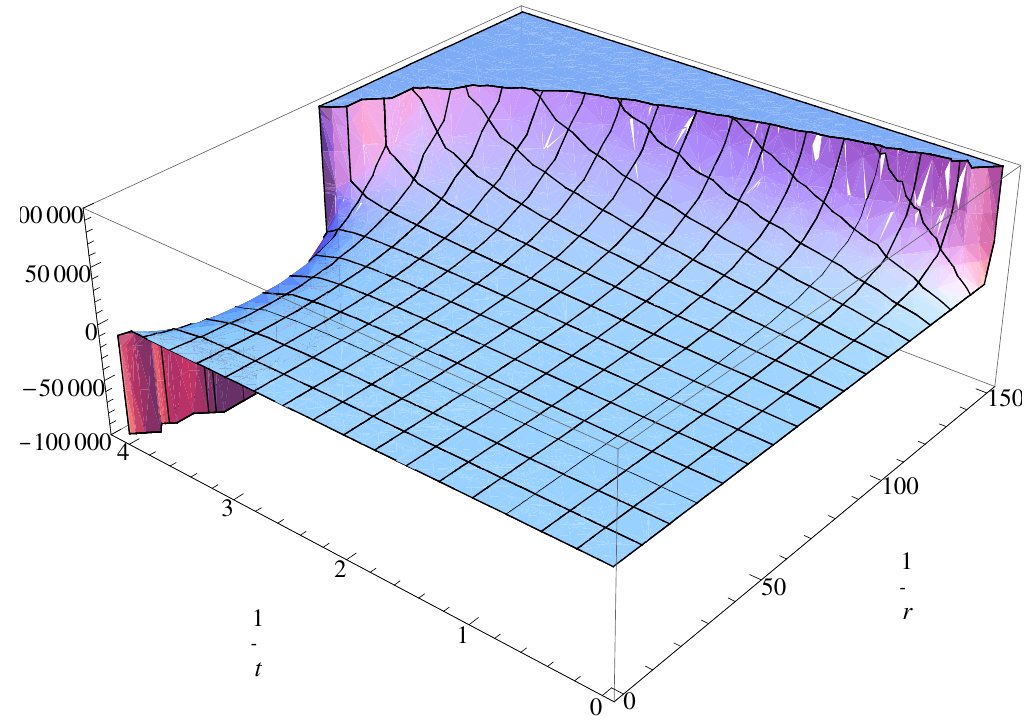}
\caption{The potential $V(r,t)$ of \eqref{potallzero2} at small values of $r$ and $t$ on the left hand side, and at large values of $r$ and $t$ (in terms of the inverse variables $1/r$ and $1/t$) on the right hand side. The parameters in \eqref{potallzero2} are chosen as $\tilde{h}=1,\,e=1,\,\lambda_2=1/2,\,\kappa=1$. We also have $c=-1/(96\pi)$ and $C=\zeta(3)/(2\pi^3)$.}
\label{fig:pot}
\end{center}
\end{figure}

When being expanded near $t=t_c$, the potential reads 
\begin{equation}
V=
\frac{A(\cR)}{t-t_c}+O(t-t_c)~,
\end{equation}
where the residue is given by
\begin{equation}
A(\cR)=\left[
\frac{(6C)^{2/3}e^2\lambda_2-8\kappa^{2/3}\tilde{h}^2(2c-\cR^2\lambda_2+4\cR^2\lambda_2^2)}{2(6C)^{2/3}\kappa\lambda_2(-2c+\cR^2\lambda_2)(2c-\cR^2\lambda_2+4\cR^2\lambda_2^2)}\right]~.
\end{equation}

There exist the critical value $\cR_{c}^{(2)}$ where  the residue vanishes, $A(\cR_{c}^{(2)})=0$, and the potential $V$ drastically changes its shape, as is shown in Fig.~2. We find
\begin{equation}
\cR_{c}^{(2)}=\frac{\sqrt{16c\kappa^{2/3}\tilde{h}^2-(6C)^{2/3}e^2\lambda_2}}{2\sqrt{2}\sqrt{\kappa^{2/3}\tilde{h}^2(1-4\lambda_2)\lambda_2}}~~.
\end{equation}
The potential $V$ at large values of $t$ or $\cR$ is not affected by that.

\begin{figure}[t]
\begin{center}
\includegraphics[width=8cm]{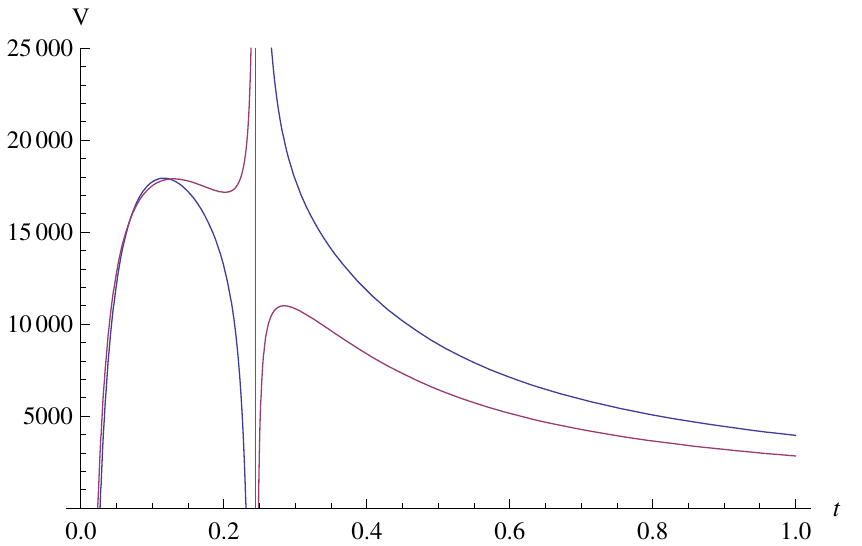}
\caption{The blue line is a section of the potential $V$ for $\cR>\cR_{c}^{(2)}$, and the purple line is a section of the potential $V$ for  $\cR<\cR_{c}^{(2)}$, at $\tilde{h}=1,\,e=1,\,\lambda_2=1/2,\,\kappa=1$. The section is taken at $r=0.013$ for the blue line and at $r=0.0145$ for the purple line.}
\label{fig:crit}
\end{center}
\end{figure}

As regards the behaviour of the potential $V$ at large values of $r$ and $t$, we find the following asymptotic expansion near
the origin in a plane $(1/r,1/t)$:

\begin{equation}
V(r,t)=
-\frac{e^2(\lambda_2-1)}{2\kappa(4\lambda_2-1)}\frac{1}{r^2}\frac{1}{t}
+\frac{3ce^2\lambda_2}{\kappa(4\lambda_2-1)}\frac{1}{r^3}\frac{1}{t}
-\frac{3\tilde{h}(\tilde{h}-2)}{2\kappa\lambda_2}\frac{1}{r}\frac{1}{t^3}
+O^5\left(\frac{1}{r},\frac{1}{t}\right).
\end{equation}

\begin{figure}[ht]
\begin{center}
\begin{minipage}[c]{6.5cm}
\includegraphics[width=6cm]{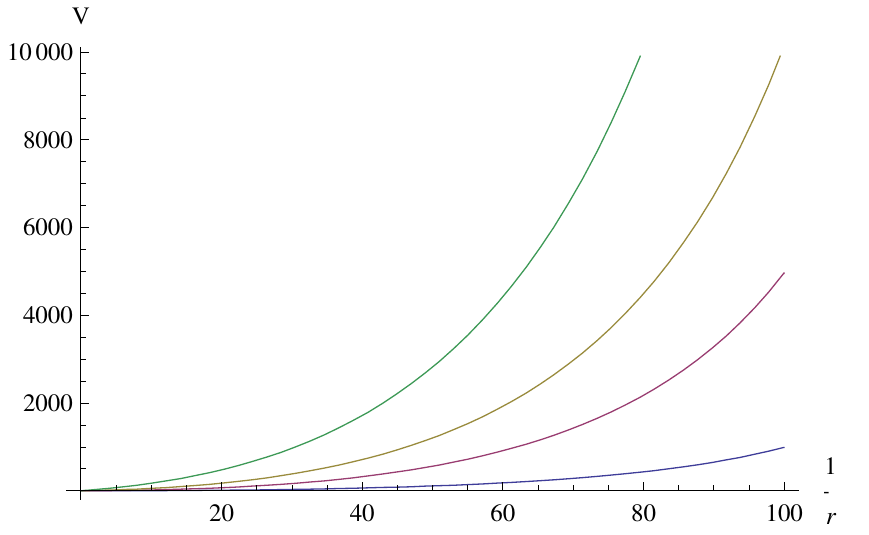}
\end{minipage}
\begin{minipage}[c]{6cm}
\includegraphics[width=6cm]{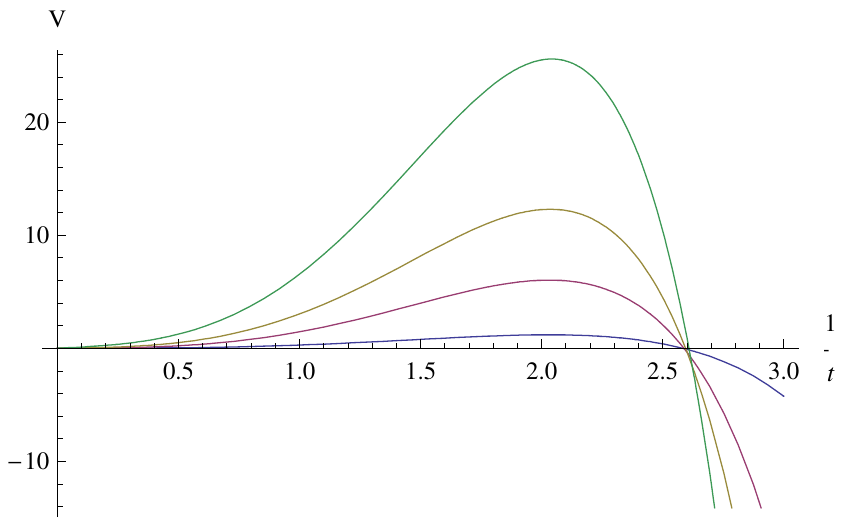}
\end{minipage}
\caption{The sliced potential $V$ at $\tilde{h}=1,\,e=1,\,\lambda_2=1/2,\,\kappa=1$. The horizontal axes show values of $1/r$ and $1/t$,
with $t=0.5$ (green), $t=1$ (olive), $t=2$ (purple)  and $t=10$ (blue), on the left hand side, and with $r=0.5$ (green), $r=1$ (olive), $r=2$ (purple) and $r=10$ (blue), on the right hand side, respectively.}
\label{fig:potential-around-infty}
\end{center}
\end{figure}

The coefficients of the three leading terms in this expansion are all positive when $1/4<\lambda_2<1$ and $\tilde{h}<2$. 
In this case, the potential $V$ takes {\it positive} values, as is shown in 
Fig.~\ref{fig:potential-around-infty}. 

We find that the (relative) first and second derivatives of the scalar potential above  are all {\it independent} upon the (flux) parameters of $V$, namely,
\begin{equation}
|V_r/V| = 1/r + O^2(1/r)~, \qquad V_{rr}/V = 2/r^2 + O^2(1/r)~, \label{twor}
\end{equation}
and
\begin{equation}
|V_t/V| =1/t + O^2(1/t)~, \qquad  V_{tt}/V = 2/t^2 + O^2(1/t)~,\label{twot}
\end{equation}
i.e. they exhibit the {\it universal} behaviour for large values of $r$ and $t$.

It follows from Eqs.~\eqref{kinpot}, \eqref{Fcl}, \eqref{Ffull} and \eqref{mett-UHMmain} in the large field approximation that the 
kinetic terms of the fields $t$ and $r$ read $3(\partial t)^2/t^2$ and $\frac{1}{4}(\partial r)^2/r^2$, respectively. In terms of the
canonical fields defined by $t=e^{\sqrt{1/6}\chi}$ and $r=e^{\sqrt{2}\varphi}$, the standard slow roll parameters are thus given
\begin{equation} \label{slowrollpar}
\varepsilon >\frac{13}{6} \qquad {\rm and} \quad \eta >2~.
\end{equation}
They violate the necessary conditions ($\varepsilon\ll 1$  and $\eta \ll 1$) for slow roll inflation. The large field limit 
also implies the runaway decompactification driving the theory towards ten dimensions, which is unacceptable 
for our Universe.

\section{Conclusion}
\label{sec-concl}

We considered a simple class of flux compactifications  of type IIA strings, preserving $N=2$ local supersymmetry in the four-dimensional low energy effective action. When back-reaction of fluxes is ignored, we obtained the non-perturbative scalar potential  that leads to the axion stabilization. The latter greatly simplifies the scalar potential as merely a function of dilaton and K\"ahler moduli. We simplified it even further by going to the perturbative approximation where all instanton contributions are ignored. We used the restricted set of electric fluxes, without magnetic fluxes and with vanishing Romans mass, which automatically obeys the tadpole cancellation condition.
 It also preserves $N=2$ local supersymmetry that allowed us to control (in principle) all quantum corrections and do explicit calculations. Even though we assumed only one K\"ahler modulus,
we expect that our results qualitatively do not change with a larger number of K\"ahler moduli and, perhaps, even with a larger number of
hypermultiplets.

We found the universal behaviour of the scalar potential for large $r$ and $t$, i.e. in the perturbative region, where all slow roll parameters become independent upon the flux parameters. 
Our results extend the applicability of the "no-go" statement \cite{Hertzberg:2007wc}, ruling out slow roll inflation in type 
IIA/CY strings, {\it beyond} the semi-classical approximation in the case of rigid CY.
The absence of viable inflation in type IIA strings on rigid CY with $N=2$ supersymmetry in four dimensions is  also related to the absence of meta-stable vacua found in \cite{Alexandrov:2016plh} for a limited range of the string coupling values, but including all D-instanton corrections.

Though unbroken $N=2$ supersymmetry certainly does not describe our universe, it may facilitate a construction of viable inflationary models 
when using $N=2$ gauged supergravity as the starting point (or as the 0th-order approximation) and then breaking extended supersymmetry
by additional structures, such as orientifolds with negative tension. However, this would require a much better understanding of quantum effects in $N=1$ supersymmetric flux compactifications, which are not under theoretical control at present.

\section*{Acknowledgements}

SVK is supported by a Grant-in-Aid of the Japanese Society for Promotion of Science (JSPS) under No.~26400252,
the World Premier International Research Centre Initiative (WPI Initiative), MEXT, Japan, and the Competitiveness Enhancement Program of Tomsk Polytechnic University in Russia. One of the authors (SVK) thanks NORDITA in Stockholm, Sweden, for kind hospitality extended to him during completion of this paper. The authors are also grateful
to the referee for careful reading of the manuscript and critical comments.

\section*{Appendix A: special geometry}
\label{subsec-special}

A special K\"ahler manifold $\cMsk$ is determined by a holomorphic prepotential $F(X^I)$, a homogeneous function of degree 2.
The homogeneous coordinates $X^I$ are related to the coordinates on the manifold $z^i$ by $z^i=X^I/X^0$ and,
for simplicity, we choose the gauge where $X^0=1$. Given the prepotential, it is convenient to define the matrix
\be
N_{IJ}=-2\Im F_{IJ}.
\label{defN}
\ee
It is invertible, but has a split signature $(b_2,1)$. A related invertible matrix with a definite signature
can be constructed as follows. Let us define
\be
\cN_{IJ}=\bF_{IJ}-\frac{\I \, N_{IK}X^K N_{JL}X^L}{N_{MN}X^M X^N}\,.
\label{defcN}
\ee
$\cN_{IJ}$ appears as the coupling matrix of the gauge fields in the low-energy effective action
and its imaginary part is negative definite.

In terms of the matrix \eqref{defN}, the K\"ahler potential on $\cMsk$ is given by
\be
\cK=-\log\(X^I N_{IJ}\bX^J\).
\label{defcK}
\ee
Its derivatives with respect to $z^i$ and $\bz^{\bi}$ are
\begin{subequations}
\bea
\cK_i&=&-e^{\cK} N_{i I}\bX^I,
\label{derK}\\
\cK_{i\bj} &=&-e^{\cK} N_{ij}+\cK_i\cK_{\bj},
\label{derKKb}
\eea
\end{subequations}
where we have used homogeneity of the holomorphic prepotential and $z^i=b^i+\I t^i$.
In particular, \eqref{derKKb} provides the metric on $\cMsk$.

\section*{Appendix B: the UH metric}
\label{subap-metric}

To write down the metric computed in \cite{Alexandrov:2014sya},  let us summarize the data characterizing a rigid CY manifold:
\begin{itemize}
\item
The intersection numbers $\kappa_{ijk}$, which specify the classical holomorphic prepotential \eqref{Fcl}
on the K\"ahler moduli space.
\item
The Euler characteristic $\chi_\CY=2h^{1,1}>0$, which appears in the $\alpha'$-corrected prepotential \eqref{Ffull},
and is always positive for rigid $\CY$. We also use the following parameter:
\be
c=-\frac{\chi_\CY}{192\pi}=-\frac{\pi^2}{48\zeta(3)}\, C.
\label{def-c}
\ee
\item
The complex number
\be
\lambda\equiv \lambda_1 -\I \lambda_2=\frac{\int_\cB\Omega}{\int_\cA\Omega}
\label{prepUHM}
\ee
given by the ratio of periods of the holomorphic 3-form $\Omega\in H^{3,0}(\CY)$
over an integral symplectic basis $(\cA,\cB)$ of $H_3(\CY,\IZ)$. The geometry requires that $\lambda_2>0$,
which explains the minus sign in \eqref{prepUHM}.
\item
The generalized Donaldson-Thomas (DT) invariants $\Omega_\gamma$,
which are integers counting, roughly, the number of BPS instantons
of charge $\gamma=(p,q)$. In the case of the vanishing magnetic charge $p$
and arbitrary electric charge $q$, they coincide with the Euler characteristic,
\be
\Omega_{(0,q)}=\chi_\CY.
\label{Omq}
\ee
\end{itemize}

The central charge
\be
Z_\gamma=q-\lambda p
\label{cencharge}
\ee
characterizes a D-instanton of charge $\gamma$. It is used to define the function
\be
\cX_\gamma(t)=(-1)^{qp} \exp\[-2\pi\I\(q\zeta-p\tzeta+\cR\(\varpi^{-1}Z_\gamma-\varpi \bZ_\gamma\)\)\],
\label{defcX}
\ee
where $\cR$ is a function on the moduli space, which is fixed below.
Geometrically, $t$ parametrizes the fiber of the twistor space $\cZ$, a $\CP$ bundle over $\cM_H$,
whereas $\cX_\gamma$ are Fourier modes of holomorphic Darboux coordinates on
$\cZ$ \cite{Alexandrov:2008nk,Alexandrov:2008gh}. Using \eqref{defcX}, one defines
\be
\begin{array}{rclrcl}
\Igg{}& = &\displaystyle
\int_{\ellg{\gamma}}\frac{\d \varpi}{\varpi}\,
\log\(1-\cX_\gamma\),
\qquad &
\rIg&=&
\displaystyle \int_{\ellg{\gamma}}\frac{\d \varpi}{\varpi}\,
\frac{\cX_\gamma}{1-\cX_\gamma}\, ,
\\
\Igpm& = &\displaystyle
\pm\int_{\ellg{\gamma}}\frac{\d \varpi}{\varpi^{1\pm 1}}\,\log\(1-\cX_\gamma\),
\qquad &
\rIgpm&=&
\displaystyle
\pm\int_{\ellg{\gamma}}\frac{\d \varpi}{\varpi^{1\pm 1}}\,
\frac{\cX_\gamma}{1-\cX_\gamma}\, ,
\end{array}
\label{def-JJJ}
\ee
where $\ellg{\gamma}$ is a contour on $\CP$ joining $t=0$ and $t=\infty$ along the direction fixed by the phase
of the central charge, $\ellg{\gamma}=\I Z_\gamma \IR^+$.
Expanding the integrands in powers of $\cX_\gamma$, these functions can be expressed as series of the modified Bessel functions of the second kind $K_n$.

The set of functions \eqref{def-JJJ} encodes the D-instanton corrections to the moduli space.
The related quantities appearing in the metric are
\begin{itemize}
\item
the functions
\be
\begin{split}
\vl =\frac{1}{2\pi}\sum_\gamma \Om{\gamma} |Z_\gamma|^2 \rIgm,
&\qquad
\Min= 2\lambda_2-\frac{1}{2\pi}\sum_\gamma \Om{\gamma} |Z_\gamma|^2 \rIg,
\\
\Uin =&\,\Min+\Min^{-1}|\vl|^2,
\end{split}
\label{Ab-UHM} 
\ee

\item
the one-forms
\bea
\hspace{-1.2cm}
\cY &=& \de\tzeta-\lambda \de\zeta-\frac{\I}{4\pi}\sum_\gamma \Om{\gamma}
Z_\gamma \(\rIg-\vl\Min^{-1}\rIgp\)\(q\de \zeta-p\de\tzeta\)
-\frac{2\I\vl}{ \cR \Min}\, \de r, \nonumber
\\
\hspace{-1.2cm}
\cVs &=&\frac{2r}{\pi\lambda_2\cR\Uin}\sum_\gamma \Om{\gamma} Z_\gamma
\(\rIgp+\bvl\Min^{-1} \rIg\)\[\(q-\lambda_1 p\)\(\de\tzeta-\lambda_1\de\zeta\)+\lambda_2^2 p\de\zeta \]. \nonumber
\label{conn-UHM}
\eea
\end{itemize}

Finally, the function $\cR$ entering \eqref{defcX} is implicitly determined as a solution to the following equation:
\be
r= \frac{\lambda_2\cR^2}{2}-c
-\frac{ \I\cR}{32\pi^2}\sum\limits_{\gamma}\hng{} \(Z_\gamma\Igp+\bZ_\gamma\Igm\),
\label{r-UHM}
\ee
where $r=e^\phi$ is the four-dimensional dilaton.

With all the notations above, the D-instanton corrected metric on the four-dimensional hypermultiplet moduli space is given by
Eq.~(\ref{mett-UHMmain}) in the main text. As was proven in \cite{Alexandrov:2014sya} , the metric \eqref{mett-UHMmain}
agrees with the Tod ansatz \cite{MR1423177} where the role of the Tod potential satisfying the Toda equation
is played by the function $T=2\log(\cR/2)$. 

\providecommand{\href}[2]{#2}\begingroup\raggedright\endgroup

\end{document}